\documentclass[preprint, aps, amsmath,showkeys, nofootinbib, superscriptaddress]{revtex4-2}

\usepackage{ulem}
\usepackage{graphicx} 
\usepackage{color}
\usepackage{epstopdf}
\usepackage{multirow}
\usepackage{bm}
 
\begin{document} 

\title{\textbf{First application of a microscopic $K^-NN$ absorption model in calculations of kaonic atoms}} 

\author{J.~\'{O}bertov\'{a} }
\email{jaroslava.obertova@fjfi.cvut.cz} 
\affiliation{Nuclear Physics Institute of the Czech Academy of Sciences, 25068 \v{R}e\v{z}, Czech Republic} 
\affiliation{Faculty of Nuclear Sciences and Physical Engineering, Czech Technical University in Prague, B\v{r}ehov\'{a} 7, 11519 Prague, Czech Republic}

\author{E. Friedman}
\affiliation{Racah Institute of Physics, The Hebrew University, Jerusalem 91904, Israel}

\author{J. Mare\v{s}}
\affiliation{Nuclear Physics Institute of the Czech Academy of Sciences, 25068 \v{R}e\v{z}, Czech Republic} 

\date{\today} 

\begin{abstract} 
Strong interaction energy shifts and widths in kaonic atoms are calculated for the first time using microscopic $K^- N$ + $K^- NN$ potentials derived 
from $K^- N$ scattering amplitudes constructed within SU(3) chiral coupled-channels models of meson-baryon interactions. 
The in-medium modifications of the free-space amplitudes due to the Pauli correlations are taken into account.  The $K^-N+K^-NN$ potentials evaluated for 23 nuclear species are confronted with kaonic atoms data. The description of the data significantly improves when the $K^-NN$ absorption is included. 
To get $\chi^2$ as low as for the $K^-N+$phenomenological multi-nucleon potential an additional phenomenological term, accounting for $K^{-}-3N(4N)$ processes, is still needed. However, density dependence of this phenomenological term points out  some deficiencies in the microscopic potentials and further improvements of the applied model are thus desirable. The calculated branching ratios for $K^-N$ and $K^-NN$ absorption channels in the $^{12}$C$+K^-$ atom are in reasonable agreement with the old bubble chamber data, as well as with the latest data from the AMADEUS Collaboration.
\end{abstract} 
\maketitle 

\section{Introduction}
\label{Intro}

The $K^- N$ interaction near threshold has recently been described in the framework of SU(3) chiral coupled-channels models of meson-baryon interactions. Above threshold, the models are tuned to reproduce low-energy $K^- p$ scattering and reaction data \cite{kp_crosssection1, kp_crosssection2, kp_crosssection3}. At threshold, important constraints on the $K^- p$ interaction are provided by three known threshold branching ratios \cite{kp_ratios1, kp_ratios2}, and particularly by precisely measured strong-interaction level shift and width of the $1s$ state in the kaonic hydrogen by the SIDDHARTA Collaboration~\cite{SIDDHARTA}. On the other hand,  
the $K^- n$ interaction is poorly determined due to the absence of sufficiently accurate data. 
Moreover, the $K^- N$ interaction is known substantially less below threshold, where  chiral models differ considerably in their predictions.  
Information about the subthreshold $K^-$ interaction is provided by the analyses of $\pi\Sigma$ spectra in the region of the $\Lambda(1405)$ resonance \cite{te73, he85,leps03,leps08,zy08,hades12,clas13}, dynamically generated in the SU(3) chiral coupled-channels models, and particularly by kaonic atom data throughout the periodic table.  The database of 65 points involves strong-interaction energy shifts, widths and yields (upper level widths) from CERN, Argonne, RAL, and BNL~(see \cite{fgbNPA94} and references therein). 

The above chiral models of the $K^- N$ interaction include only the $K^- N \rightarrow \pi Y\;\; (Y =\Lambda, \; \Sigma)$ decay channel. However, in the nuclear medium, $K^-$ interactions with two and more nucleons should be included as well, e.g., $K^- N N \rightarrow Y N$. In fact, the $K^-$ absorption on two and more nucleons 
amounts to about 20\% of the total $K^-$ absorption at the surface of atomic nuclei and the role of multi-nucleon absorption increases rapidly with density. 
The multi-nucleon absorption ratios were first measured in bubble chamber experiments for $K^-$ capture on a mixture of C, F, Br \cite{bubble1}, on Ne \cite{bubble2}, and C \cite{bubble3}. The $K^-$ two-nucleon absorption fractions for all possible final states on $^4$He were published in Ref.~\cite{katzPRD70}. More recently, $K^-$ three- and four-nucleon absorption fractions on $^4$He for channels with $\Lambda$ in the final state were measured in the E549 experiment at KEK \cite{KEK1}, and the FINUDA Collaboration studied the $\Sigma^- p$ emission rate in reactions of low-energy $K^-$ with light nuclei~\cite{finuda15}. Finally, the AMADEUS Collaboration reported the measured $K^-$ two-nucleon branching ratios with 
$\Lambda p$ and $\Sigma^0 p$ in the final state for low-energy antikaons absorbed in $^{12}$C \cite{amadeus16, amadeus19}. The above experiments provided valuable information 
on the $K^-$ multi-nucleon absorption in the nuclear medium. 

A recent study of kaonic atoms performed by Friedman and Gal \cite{fgNPA17} revealed that $K^-$ optical potentials based on the $K^- N$ scattering amplitudes derived within SU(3) chiral models fail to reproduce experimental data. However, once a phenomenological optical potential accounting for the $K^-$ multi-nucleon processes in nuclear matter was added, a very good global fit of kaonic atoms was achieved. Moreover, when extra constraint to reproduce simultaneously the $K^-$ single-nucleon absorption fractions from bubble chamber experiments \cite{bubble1, bubble2, bubble3} was applied, only the Kyoto-Munich \cite{km}, Prague \cite{pnlo}, and Barcelona \cite{bcn} models were found acceptable.  

The Prague and Kyoto-Munich models supplemented by the phenomenological $K^-$ multi-nucleon potential were applied in calculations of $K^-$ nuclear quasi-bound states~\cite{hmplb, hmprc}. The $K^-$ multi-nucleon absorption was found to contribute considerably to the total widths of these $K^-$ nuclear states, which then substantially exceeded the corresponding binding energies. However, since analyses  of Friedman and Gal \cite{fgNPA17} have shown that kaonic atom data probe reliably the real part of the $K^-$ optical potential up to $\approx$ 30\% and its imaginary part up to $\approx$ 50\% of nuclear density $\rho_0$, the shape of the phenomenological $K^-$ multi-nucleon potential in the nuclear interior is a matter of extrapolation to higher densities. Therefore, a proper microscopic model for $K^-$ absorption on two and more nucleons in nuclear matter is needed for a reliable description of $K^-$ absorption in atomic nuclei.

Sekihara et al. \cite{sjPRC12} developed a microscopic model for the $K^- NN$ absorption in nuclear matter employing a chiral unitary approach to a free-space $\bar{K}N$ interaction and evaluated the branching ratios of mesonic and nonmesonic $K^-$ absorption in the nuclear medium. 

Inspired by the evaluation of an $\eta'$-nucleus optical potential including  $2N$ absorption using an $\eta'$ self-energy constructed within a meson exchange formalism \cite{nagahiroPLB12}, Hrt\'{a}nkov\'{a} and Ramos \cite{hrPRC20} developed a microscopic model for the $K^- NN$ absorption in symmetric nuclear matter. 
The absorption was described within a meson-exchange picture and the primary $K^- N$ interaction strength was derived from chiral interaction models. The medium  modification of the $K^- N$ scattering amplitudes due to the Pauli correlations was taken into account, which appeared crucial. 
The derived $K^- N$ and $K^- NN$ optical potentials were applied in calculations of the $K^-$ single- and two-nucleon absorption fractions and branching ratios for various mesonic and nonmesonic channels.  

In the present work, we apply for the first time the microscopic $K^-N+K^-NN$ potentials derived from chiral $K^-N$ scattering amplitudes in calculations of the strong interaction energy shifts and widths in kaonic atoms.  The chiral amplitudes are constructed within the Barcelona and Prague models and the in-medium modifications of the free-space amplitudes due to the Pauli principle are taken into account. In addition, we calculate branching ratios for all $K^- N$ and $K^- NN$ absorption channels in the $^{12}$C+$K^-$ atom. 

The paper is organized as follows. Section II provides a brief description of the formalism used to derive the microscopic $K^-N$ and $K^-NN$ optical potentials, followed by a construction of underlying in-medium $K^-N$ amplitudes from free-space amplitudes obtained within chiral interaction models. A discussion of subthreshold kinematics applied to kaonic atoms is also presented. In Section III, the microscopic 
$K^-N$ and $K^-NN$ optical potentials are confronted with kaonic atoms data. It is demonstrated that the description of the data improves considerably when the $K^-NN$ potential is included. An additional phenomenological term, introduced to incorporate missing $K^- -3N$($4N$) processes, is discussed. Finally, 
branching ratios for all $K^- N$ and $K^- NN$ absorption channels in  $^{12}$C+$K^-$, calculated using microscopic $K^- N + K^- NN$ potentials are compared with old bubble chamber data and with the branching ratios reported recently  by the AMADEUS Collaboration. 
A brief summary is given in Section IV. 

\section{Model}
\label{model}
 
This section provides a brief introduction to the kaonic atom methodology 
and to our microscopic $K^- NN$ absorption model applied in the present calculations. 
For more details see Refs.~\cite{fgNPA17,hrPRC20}.
 
The binding energies $B_{K^-}$ and widths $\Gamma_{K^-}$ of $K^-$ atomic states are determined by solution of the Klein-Gordon equation 
\begin{equation}\label{KG}
 \left[ \vec{\nabla}^2  + \tilde{\omega}_{K^-}^2 -m_{K^-}^2 -\Pi_{K^-}(\omega_{K^-},\rho) \right]\phi_{K^-} = 0~,
\end{equation}
where $\tilde{\omega}_{K^-} = m_{K^-} - B_{K^-} -{\rm i}\Gamma_{K^-}/2 -V_C= \omega_{K^-} - V_C$,  $m_{K^-}$ is the $K^-$ mass, $\omega_{K^-}$ stands for a complex kaon energy, $V_C$ is 
the Coulomb potential introduced via the minimal substitution~\cite{kkwPRL90}, and  $\rho$ is the nuclear density distribution. The $K^-$ interaction with the nuclear medium is described by the energy- and density-dependent kaon self-energy operator
\begin{equation}\label{piK}
\Pi_{K^-} = 2 {\mu}_{K^-}(V_{K^-N}+V_{K^-NN})~,
\end{equation}
where $V_{K^-N}$ denotes the $K^-$ single-nucleon potential, $V_{K^-NN}$ is the $K^-$ two-nucleon potential, and $\mu_{K^-}$ is the $K^-$-nucleus reduced mass.
The $K^-N$ potential is taken in a $t\rho$ form
\begin{equation}\label{V_KN}
2\mu_{K^-}V_{K^-N}=-4\pi \left(1+\frac{A-1}{A}\frac{\mu_{K^-}}{m_N} \right)\left(F_0\frac{1}{2}\rho_p + F_1\left(\frac{1}{2}\rho_p+\rho_n\right)\right)~,
\end{equation}
where $F_0$ and $F_1$ are the isospin 0 and 1 $s$-wave in-medium amplitudes in the $K^-N\rightarrow K^-N$ channel, respectively, and $m_N$ is the nucleon mass. The symbols $\rho_p$ and $\rho_n$ denote proton and neutron density distributions, respectively, calculated within the relativistic mean field model TM2 for light and medium mass nuclei ($A<40$) and TM1 for heavier nuclei ($A\geq40$)~\cite{Toki}. It is to be noted that very similar results were obtained for other density distributions (two-parameter Fermi distributions and the NL-SH parametrization \cite{nlsh}).

\begin{figure}[t!]
\includegraphics[width=0.2\textwidth]{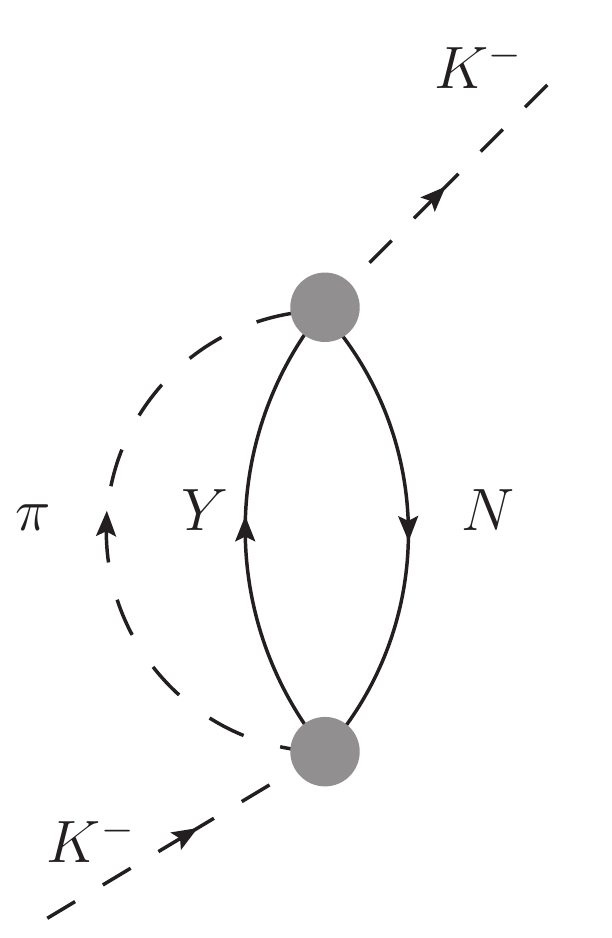}
\caption{\label{fig:one_loop_diagrams} Feynman diagram for $K^-$ single-nucleon absorption in nuclear matter. The shaded circles denote the $K^-N\rightarrow \pi Y,~(Y=\Lambda,~\Sigma)$ t-matrices derived from a chiral coupled-channels meson-baryon interaction model. Figure adapted from Ref.~\cite{hrPRC20}.}
\end{figure}

Apart from that, the imaginary part of the $K^-$ single-nucleon potential can be evaluated as the self-energy of the Feynman diagram shown in Fig.~\ref{fig:one_loop_diagrams} \cite{hrPRC20}. Here, the shaded circles denote the $K^-N\rightarrow \pi Y,~(Y=\Lambda,~\Sigma)$ t-matrices derived from a chiral coupled-channels meson-baryon interaction model. The total imaginary $K^-N$ potential is then built as a sum of the contributions from each absorption channel listed on the left-hand side of Table~\ref{tab:channels},
\begin{equation}\label{eq:imVkn}
 \text{Im}V_{K^-N}=\sum_{\rm channels} \text{Im}V_{K^-N\rightarrow \pi Y}~.
\end{equation}
We checked that \text{Im}V$_{K^-N}$ from Eq.~\eqref{eq:imVkn} and the imaginary part of the '$t\rho$' potential  [Eq.~\eqref{V_KN}] yield identical results for kaonic atoms.

\begin{table}[b!]
\caption{Channels considered for $K^-$ single-nucleon (left) and two-nucleon (right) absorption in nuclear matter. }
 \begin{tabular}{cl|cl} \label{tab:channels}
  $K^-N$ & $\rightarrow \pi Y$ & $K^-N_1N_2$  &$\rightarrow YN$ \\ \hline
   $K^-p$& $\rightarrow \pi^0 \Lambda$ & $K^-pp$ & $\rightarrow \Lambda p$ \\
   & $\rightarrow \pi^0 \Sigma^0$ & & $\rightarrow \Sigma^0 p$ \\
   & $\rightarrow \pi^+ \Sigma^-$ & & $\rightarrow \Sigma^+ n$  \\
  & $\rightarrow \pi^- \Sigma^+$ & $K^-pn(np)$ & $\rightarrow \Lambda n$ \\ 
  $K^-n$ & $\rightarrow \pi^- \Lambda$  & & $\rightarrow \Sigma^0 n$ \\ 
  & $\rightarrow \pi^- \Sigma^0$ & & $\rightarrow \Sigma^- p$ \\
  & $\rightarrow \pi^0 \Sigma^-$ & $K^-nn$ & $\rightarrow \Sigma^- n $ \\ 
 \end{tabular}
\end{table}

\begin{figure}[t!]
\includegraphics[width=0.7\textwidth]{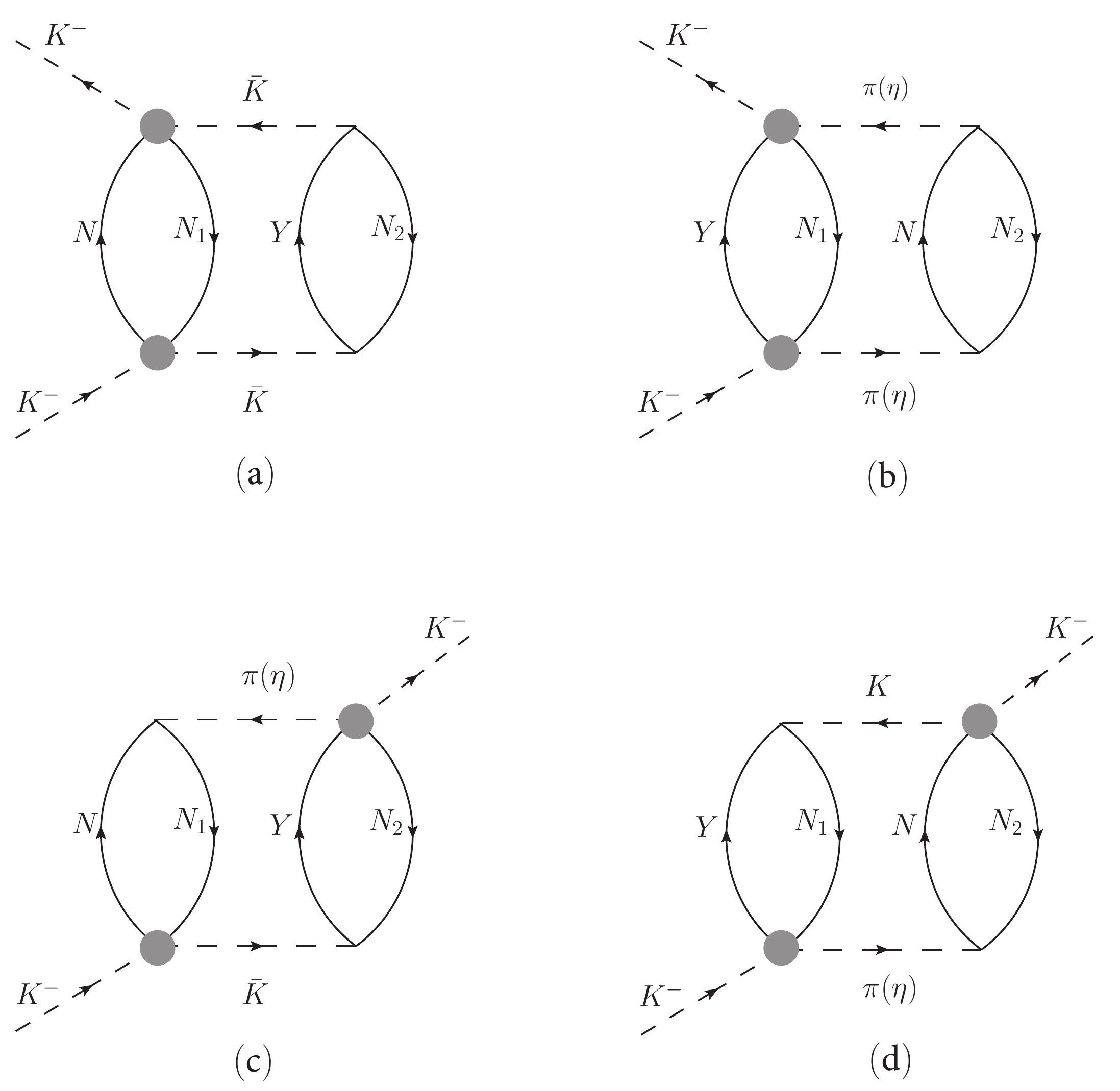}
\caption{\label{fig:direct_diagrams} Direct Feynman diagrams for $K^-$ absorption on two nucleons $N_1,~N_2$ in nuclear matter. The shaded circles denote the $K^-N$ t-matrices derived from a chiral coupled-channels meson-baryon interaction model. Figure adapted from Ref.~\cite{hrPRC20}.}
\end{figure}

\begin{figure}[t!]
\includegraphics[width=0.7\textwidth]{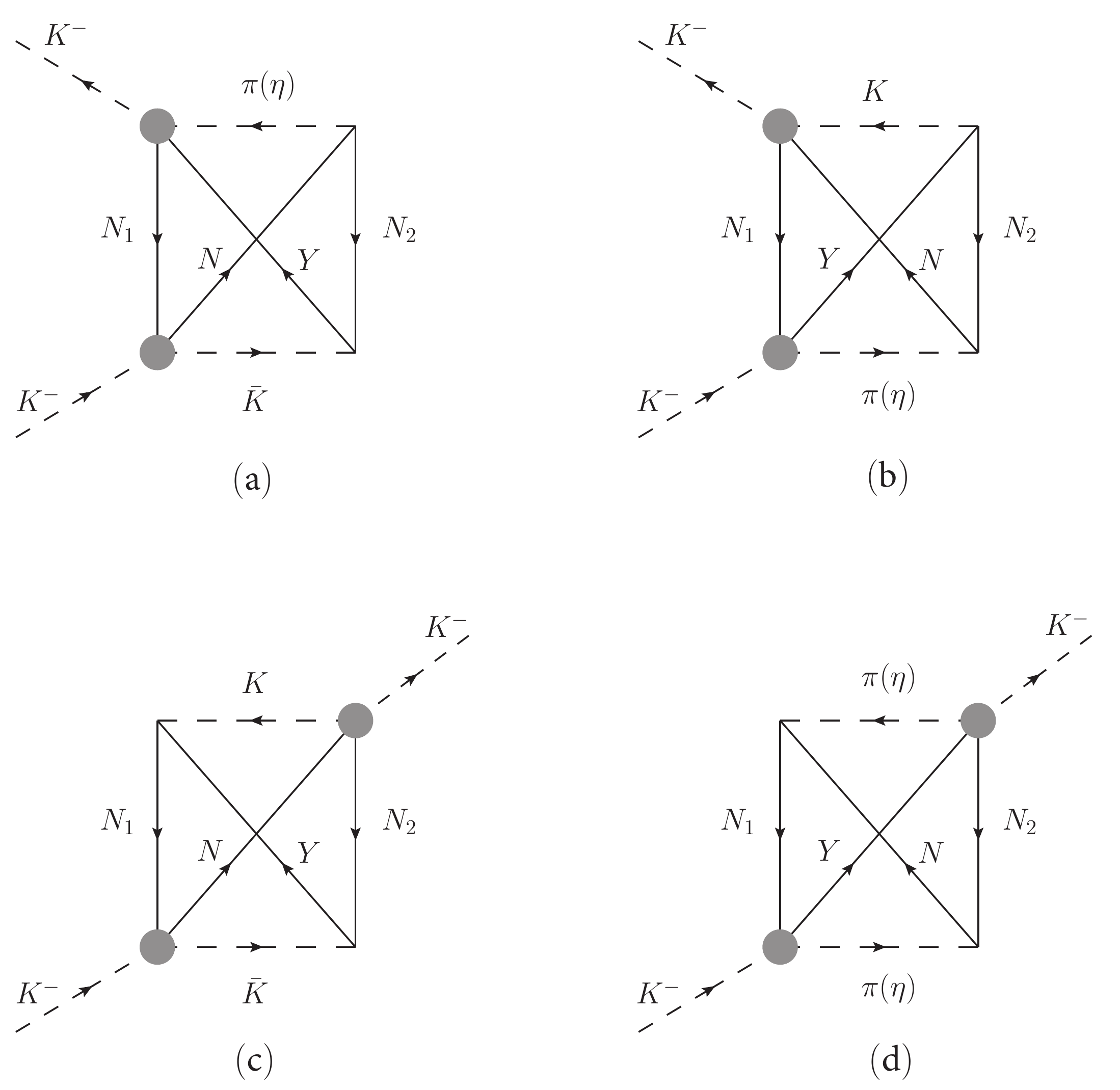}
\caption{\label{fig:crossed_diagrams} Exchange Feynman diagrams for $K^-$ absorption on two nucleons $N_1,~N_2$ in nuclear matter. The shaded circles denote the $K^-N$ t-matrices derived from a chiral coupled-channels meson-baryon interaction model. Figure adapted from Ref.~\cite{hrPRC20}.}
\end{figure}

The $K^-NN$ potential is constructed within our recently developed microscopic model~\cite{hrPRC20}. The $K^-NN$ absorption is described as a process with different intermediate virtual mesons exchanged ($\overline{K},~\pi,~\eta$) as represented by Feynman diagrams in Figs.~\ref{fig:direct_diagrams} and \ref{fig:crossed_diagrams}. Diagrams in Fig.~\ref{fig:crossed_diagrams} are obtained by antisymmetrizing the initial $N_1N_2$ system and exchanging the $N$ and $Y$ lines in the final state. The channels considered for $K^-NN$ absorption in nuclear matter are listed on the right-hand side of Table~\ref{tab:channels}. Each channel can proceed via direct (Fig.~\ref{fig:direct_diagrams}) and exchange (Fig.~\ref{fig:crossed_diagrams}) diagrams with the corresponding intermediate mesons. 

The total $K^-NN$ potential is obtained as a sum of contributions coming from the direct and exchange diagrams for all considered channels
\begin{equation}
    V_{K^-NN}=\sum_{\rm channels} V_{K^-NN}^{\rm direct} + V_{K^-NN}^{\rm exchange}~.
\end{equation}

The underlying chiral $K^-N$ scattering amplitudes are derived within the Barcelona (BCN) \cite{bcn} and Prague (P) \cite{pnlo} models. These models supplemented by a phenomenological multinucleon term were found to describe simultaneously the $K^-$ atoms data and the $K^-$ single-nucleon absorption fraction \cite{fgNPA17}. In the nuclear medium, the $K^-N$ interaction is modified due to the Pauli principle~\cite{koch94, wkw96} and hadron self-energies ~\cite{lutz98, ro00, cfgm01, cfggm11}. We incorporated the in-medium modifications due to the Pauli blocking using two different approaches in this work. First, the Pauli blocking effect was included directly in the chiral amplitudes by restricting the nucleon momentum in the intermediate meson-nucleon loops of the uniterized amplitude to be larger than the Fermi momentum (denoted further by `Pauli').
The second method is based on the multiple scattering approach by T. Wass, M. Rho and W. Weise (denoted further by 'WRW')~\cite{wrw}. Here, the in-medium isospin 0 and 1 amplitudes $F_0$ and $F_1$, respectively, for the diagonal channels $K^-N \rightarrow K^-N$ are evaluated from the free-space amplitudes $f_{K^-N\rightarrow K^-N}$ using the following formulas:
\begin{equation}\label{Eq.:in-med amp 11}
F_{1}=\frac{f_{K^-n\rightarrow K^-n}(\sqrt{s})}{1+\frac{1}{4}\xi_k c_{lab} f_{K^-n\rightarrow K^-n}(\sqrt{s}) \rho}~, \quad F_{0}=\frac{[2f_{K^-p\rightarrow K^-p}(\sqrt{s})-f_{K^-n\rightarrow K^-n}(\sqrt{s})]}{1+\frac{1}{4}\xi_k c_{lab}[2f_{K^-p\rightarrow K^-p}(\sqrt{s}) - f_{K^-n\rightarrow K^-n}(\sqrt{s})] \rho}~.
\end{equation}
Amplitudes for the non-diagonal channels $K^-N\rightarrow \pi/\eta Y~ (Y=\Lambda, \Sigma)$ are modified as follows
\begin{equation}\label{Eq.:in-med amp 12}
    F_{1(0)}^{\pi/\eta Y} = \frac{f_{1(0)}^{\pi/\eta Y}(\sqrt{s})}{1+\frac{1}{4}\xi_k c_{\rm lab} f_{1(0)}(\sqrt{s}) \rho}
\end{equation}
where $f_{1(0)}^{\pi/\eta Y}$ is the free-space isospin 0 or 1 amplitude in the channel $K^-N\rightarrow \pi/\eta Y$, and $f_{1(0)}$ denotes corresponding free-space isospin amplitude in the channel $K^-N\rightarrow K^-N$. Here,
\begin{equation}
c_{\rm lab}=1+\frac{A-1}{A}\frac{\mu_{K^-}}{m_N}
\end{equation} 
and
\begin{equation}\label{ksi}
 \xi_k=\frac{9\pi}{p_{\rm F}^2}\,4 I,\;\;\;\;\; I = \int_0^{\infty} \frac{dr}{r} \exp(ikr)j_1^2(p_Fr), 
\end{equation}
$p_{\rm F}$ is the Fermi momentum corresponding to density $\rho = 2p_{\rm F}^3/(3\pi^2)$, $j_1$ is the spherical Bessel function and $k$ is the kaon momentum
\begin{equation}\label{eq:k_wrw}
k= \sqrt{\omega_{K^-}^2-m_{K^-}^2}~.
\end{equation}
The integral $I$ in Eq.~\eqref{ksi} can be evaluated analytically \cite{fgNPA17}
\begin{equation}
 4 I(q) = 1- \frac{q^2}{6} + \frac{q^2}{4}\left(2+ \frac{q^2}{6}\right) \ln \left(1+\frac{4}{q^2}\right) - \frac{4}{3}q\left( \frac{\pi}{2} - \arctan(q/2)\right)~,
\end{equation}
where $q=-ik/p_F$.
\begin{figure}[b]
\begin{center}
\includegraphics[width=0.48\textwidth]{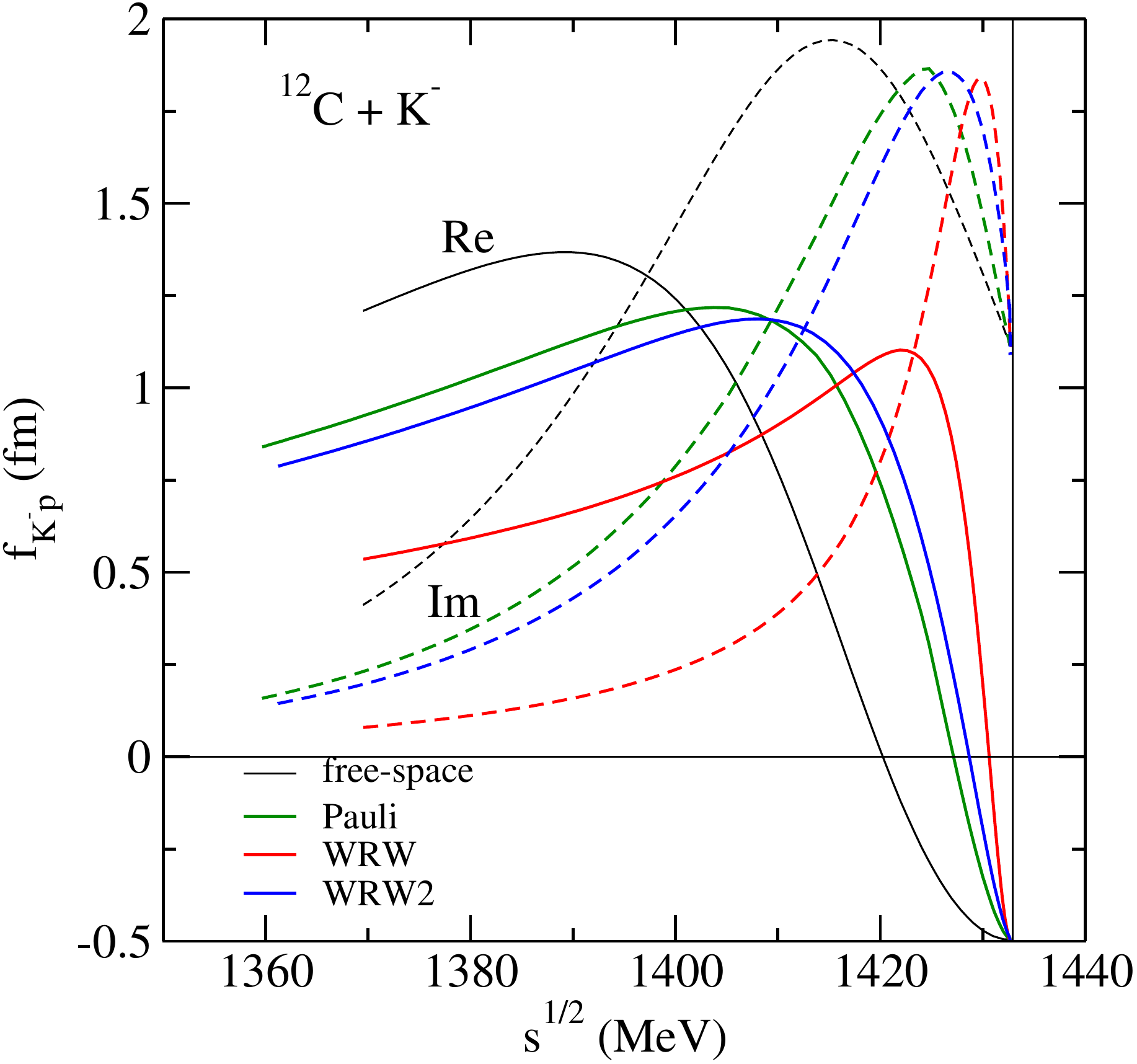} \hspace{10pt}
\includegraphics[width=0.47\textwidth]{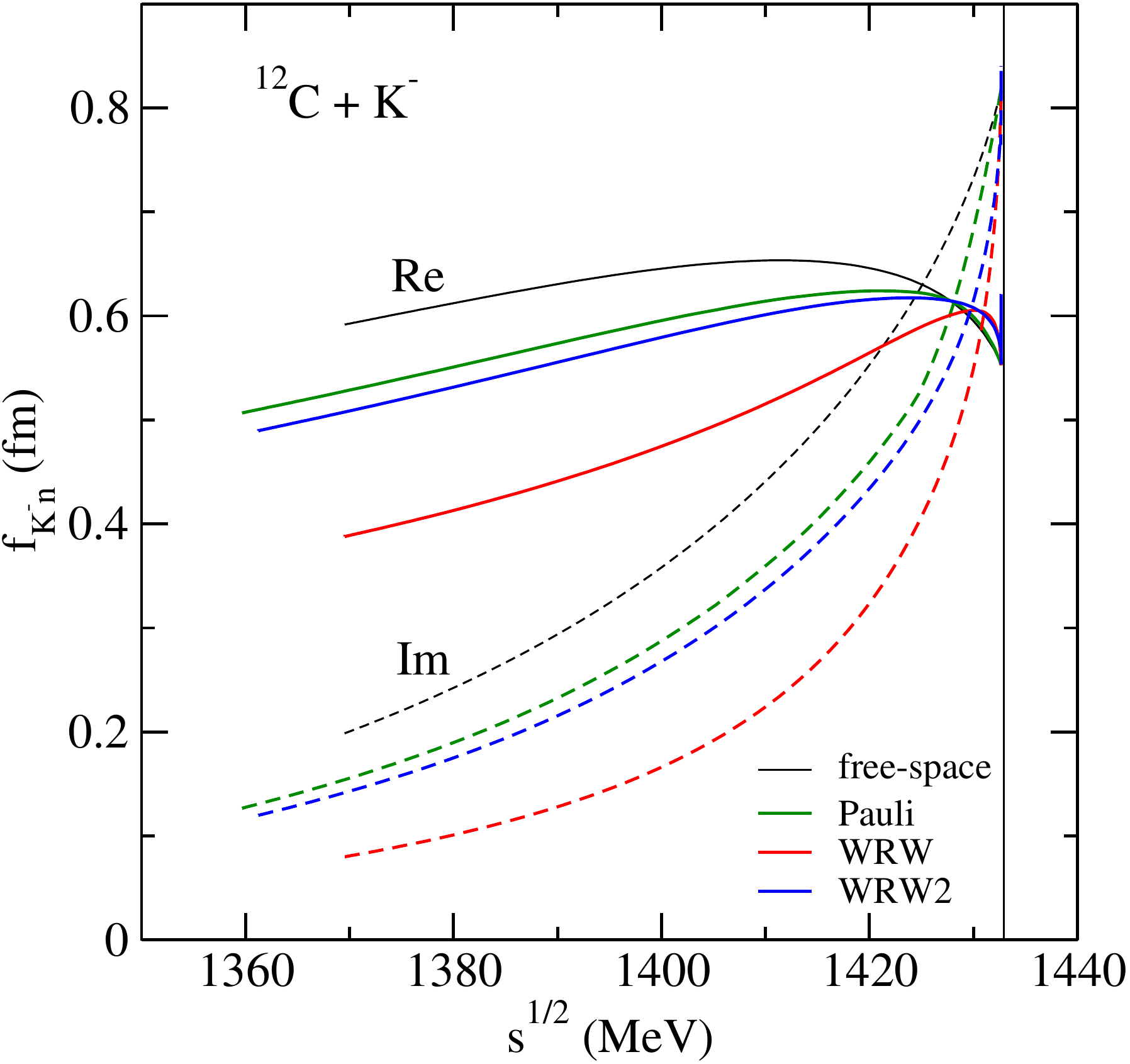}
\end{center}
\caption{\label{fig:Pauli_WRW_amp} Comparison of the Pauli (green) and WRW (red) modified $K^-p$ (left) and $K^-n$ (right) BCN amplitudes, printed from the $^{12}$C$+K^-$ atom calculation with the $K^-N+K^-NN$ potential. 'WRW2' (blue) denotes WRW amplitudes evaluated using a cms kaon momentum $k_{\rm cms}$ (see text for details). The vertical line denotes the $K^-N$ threshold.}
\end{figure}

In Fig.~\ref{fig:Pauli_WRW_amp}, we compare the Pauli (green) and WRW (red) modified $K^-p$ and $K^-n$ BCN amplitudes in the $^{12}$C$+K^-$ atom, calculated with the $K^-N+K^-NN$ potential as functions of $\sqrt{s}$, where $s$ is the Mandelstam variable. The free-space amplitudes (black) are also shown for comparison. The in-medium modification causes the peak of the $K^-p$ amplitude to shift towards higher $\sqrt{s}$. There is a noticeable difference between the Pauli and WRW modified $K^-p$ (left panel) and $K^-n$ (right panel) amplitudes. The WRW method yields smaller values of the real and imaginary parts of the $K^-p$ and $K^-n$ amplitudes than the Pauli approach in most of the subthreshold region. The reason for the discrepancy between the Pauli and WRW amplitudes lies in the definition of momentum $k$ employed in the two methods. While in the WRW method the ansatz for $k$ is given by Eq.~\eqref{eq:k_wrw}, in the evaluation of the Pauli blocked amplitudes a center-of-mass system (cms) formula is used
\begin{equation}\label{eq:k_cms}
k_{\rm cms}=\sqrt{\frac{[s-(m_N+m_{K^-})^2][s-(m_N-m_{K^-})^2]}{4s}}~.
\end{equation}

If the expression for $k$ from Eq.~\eqref{eq:k_cms} is used in the WRW method then the resulting in-medium amplitudes get considerably closer to the Pauli amplitudes (see blue lines in Fig.~\ref{fig:Pauli_WRW_amp} denoted by 'WRW2'). Nonetheless, we used the WRW method with $k$ defined by Eq.~\eqref{eq:k_wrw} in the present calculations, following previous kaonic atoms studies \cite{fgNPA17, fgNPA}.

The in-medium amplitudes in Eqs.~\eqref{Eq.:in-med amp 11} and~\eqref{Eq.:in-med amp 12} (as well as the Pauli amplitudes) are functions of energy $\sqrt{s}$ given by the Mandelstam variable 
\begin{equation}\label{eq:s}
 s=(E_N+E_{K^-})^2-(\vec{p}_N+\vec{p}_{K^-})^2~,
\end{equation}
where $E_N=m_N-B_N$, $E_{K^-}=m_{K^-}-B_{K^-}$ and $\vec{p}_{N(K^-)}$ is the nucleon (kaon) momentum. The momentum dependent term $(\vec{p}_N+\vec{p}_{K^-})^2 
\neq 0$ in the $K^-$-nucleus cm frame and generates additional substantial downward energy 
shift \cite{cfggm11}. 
The $K^-N$ amplitudes can then be expressed as a function of energy $ \sqrt{s} = E_{\rm th} + \delta \sqrt{s}$ where $E_{\rm th}=m_N + m_{K^-}$. The relative energy $\delta \sqrt{s}$ is expanded near threshold in terms of binding and
kinetic energies (to leading order) and specific forms of density dependence are introduced ensuring that $\delta \sqrt{s} \rightarrow 0$ as $\rho \rightarrow 0$ (for details see Refs.~\cite{cfggm11,fgNPA17,hmplb, hmprc}):
\begin{equation} \label{Eq.:deltaEsLDL}
 \delta \sqrt{s}=  -B_N\frac{\rho}{\bar{\rho}}\, - \beta_N\! \left[B_{K^-}\frac{\rho}{\rho_{\rm max}} + T_N\left(\frac{\rho}{\bar{\rho}}\right)^{2/3}\!\!\!\! +V_C\left(\frac{\rho}{\rho_{\rm max}}\right)^{1/3}\right] + \beta_{K^-} {\rm Re}V_{K^-}(r)~,
\end{equation}
where $\beta_{N(K^-)}={m_{N(K^-)}}/(m_N+m_{K^-})$, $B_N=8.5$~MeV is the average binding energy per nucleon, $\rho_{\rm max}$ and $\bar{\rho}$ are the maximal and average value of the nuclear density, respectively. Since $\delta \sqrt{s}$ depends on Re$V_{K^-}$ [ and thus $f_{K^-N}(\sqrt{s})$] and $B_{K^-}$ which by themselves depend on $\sqrt{s}$, it is clear that for a given value of $B_{K^-}$, $f_{K^-N}(\sqrt{s})$
has to be determined self-consistently by iterations.

\section{Results}
\label{results}
Using the method described in Section \ref{model}, we performed calculations for 23 atomic species from lithium up to uranium. For the first time, we evaluated energy shifts and widths for lower and upper states (65 data points) using microscopic $K^-N+K^-NN$ potentials based on chiral amplitudes and compared them with available kaonic atom data. Three cases were
considered. First, the data were compared with predictions of the $K^-$ single-nucleon potential. 
Next, the $K^-$ two nucleon potential was added to the $K^-$ single-nucleon potential ($K^-N+K^-NN$) and finally, the $K^-N+K^-NN$ potential was supplemented by an additional phenomenological term
(+phen.) in order to cover $3N(4N)$ processes not included in our microscopic model: 
\begin{equation}
    V_{\rm phen} = -4\pi B \left(\frac{\rho}{\rho_0}\right)^{\alpha} \rho~,
\end{equation}
where $B$ is a complex amplitude, $\alpha$ is a positive number and $\rho_0 = 0.17$~fm$^{-3}$ is the saturation density.

\begin{table}[b]
\caption{Values of $\chi^2(65)$ resulting from comparison of predictions of $K^-N$, $K^-N+K^-NN$, and $K^-N+K^-NN+$phen. multi-N potentials with kaonic atom data. Microscopic potentials are based on the Pauli and WRW modified BCN amplitudes. Values of the complex amplitude $B$ and parameter $\alpha$ for the additional phenomenological term are presented as well.}
\vspace{10pt}
 \begin{tabular}{c|c|c|c|c|c|cc}
 
 & $K^-N$ & $K^-N+K^-NN$ & + phen. & Re$B$ (fm) & Im$B$ (fm) & $\alpha$ \\ \hline
Pauli & 825  & 565 & 105 & -1.97(13) & -0.93(11) & 1.4\\ 
WRW & 2378 & 1123 & 116 & -0.90(9) & 0.72(10) & 0.6 \\ 
\end{tabular}
\label{Tab:fit}
\end{table} 

The results of calculations using the Pauli and WRW modified BCN amplitudes are presented in Table~\ref{Tab:fit}. The description of the data improves significantly when the $K^-NN$ absorption is taken into account. The value of $\chi^2(65)$ decreases to about one half with respect to the case of pure $K^-N$ potentials, however, it still remains considerable. This suggests that some additional processes of $3N(4N)$ absorption might be missing. When the additional phenomenological term is added to the $K^-N+K^-NN$ potentials, values of $\chi^2(65)$ further decrease to $\approx 100$, which is comparable with the best fit $K^-N$ + phenomenological multi-nucleon potential, Re$B$ = -1.3 fm, Im$B$ = 1.9 fm, $\alpha=1$, $\chi^2(65)=112.3$~\cite{jarka22}. In general, the value of $\chi^2(65)$ is much lower for the potentials based on the Pauli amplitudes than for the potentials derived from the WRW amplitudes. On the other hand, the value of Im$B$ of the additional phenomenological term is negative for the Pauli amplitudes which means negative absorption. In other words, the $K^-N+K^-NN$ potentials based on the Pauli BCN amplitudes seem to be too absorptive in the relevant density region and there is no space for additional absorption from $3N(4N)$ processes. On the contrary, the resulting best fit value of Im$B=0.72 \pm 0.10$ fm is positive for potentials based on the WRW modified amplitudes. It is about half of the value obtained in the fit with $K^-N$ + phenomenological multi-nucleon optical potential, Im$B=1.9$ fm, which seems reasonable. 

The parameter $\alpha$ controls density dependence of the additional phenomenological term. It is expected to be $\alpha \geq 2$ for the $3N(4N)$ processes. However, the fit yields values of $\alpha$ lower than 2 in both cases, which indicates certain deficiencies in the microscopic $K^-$ potentials. Although the WRW amplitudes seem to yield reasonable fit with a positive value of Im$B$, the unexpected density dependence ($\alpha<2$) of the additional phenomenological term may suggest that the $K^-N+K^-NN$ microscopic potential should be more absorptive in the relevant density region. 

One of the possible improvements is the proper inclusion of self-energy insertions in terms of hadron-nucleon potentials for the intermediate hadrons in the amplitudes employed here. The self-energy effects are expected to partly compensate for the upward Pauli shift of the scattering amplitudes~\cite{lutz98, ro00, cfgm01, cfggm11}.

\begin{figure}[t!]
\begin{center}
\includegraphics[width=0.7\textwidth]{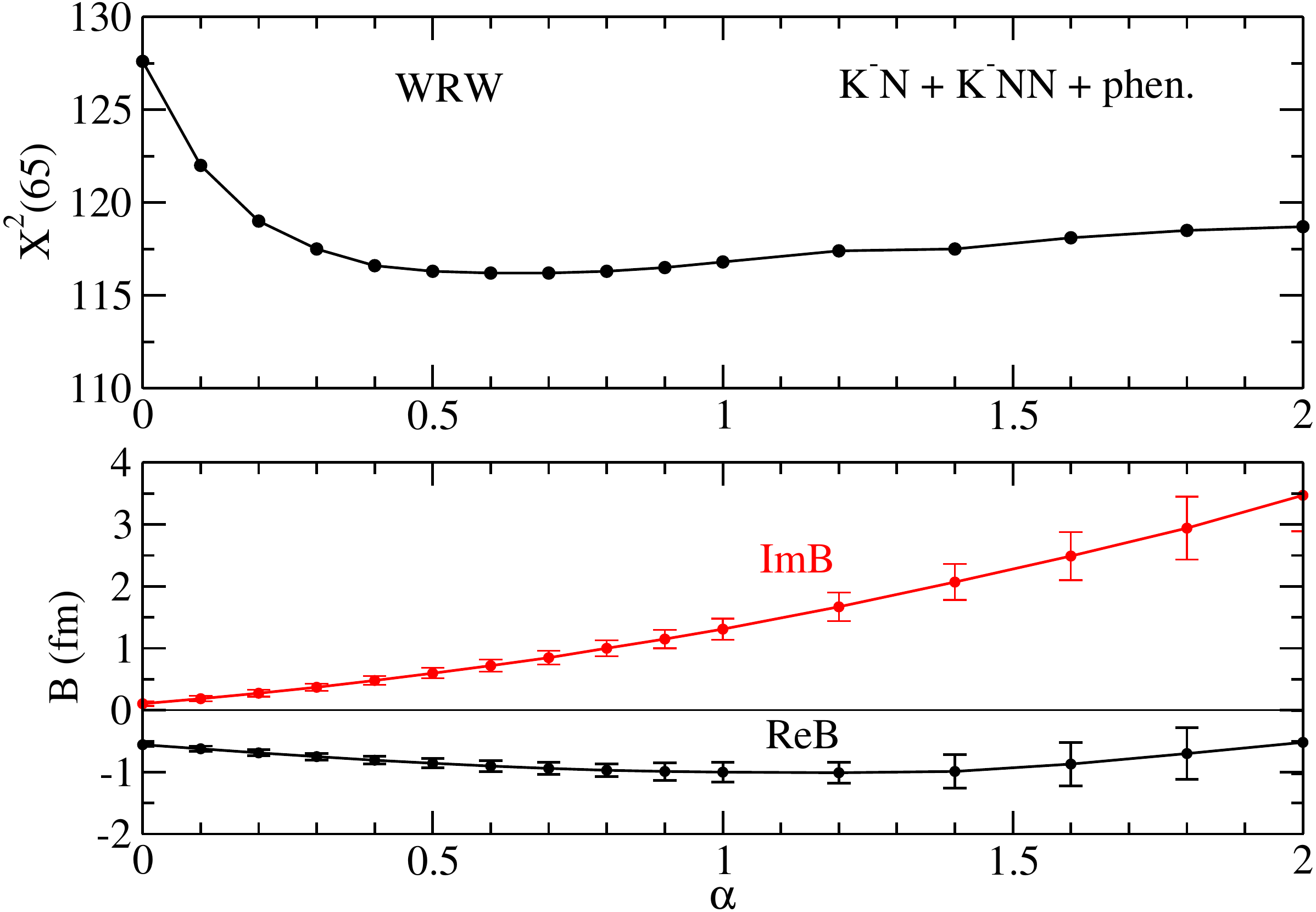}
\end{center}
\caption{Values of $\chi^2(65)$ (upper panel) and the complex amplitude $B$ (lower panel) as functions of the parameter $\alpha$ from best fits based on $K^-N+K^-NN$ potentials using the WRW BCN amplitudes.}
\label{fig:fit_WRW}
\end{figure}

\begin{figure}[b!]
\begin{center}
\includegraphics[width=0.7\textwidth]{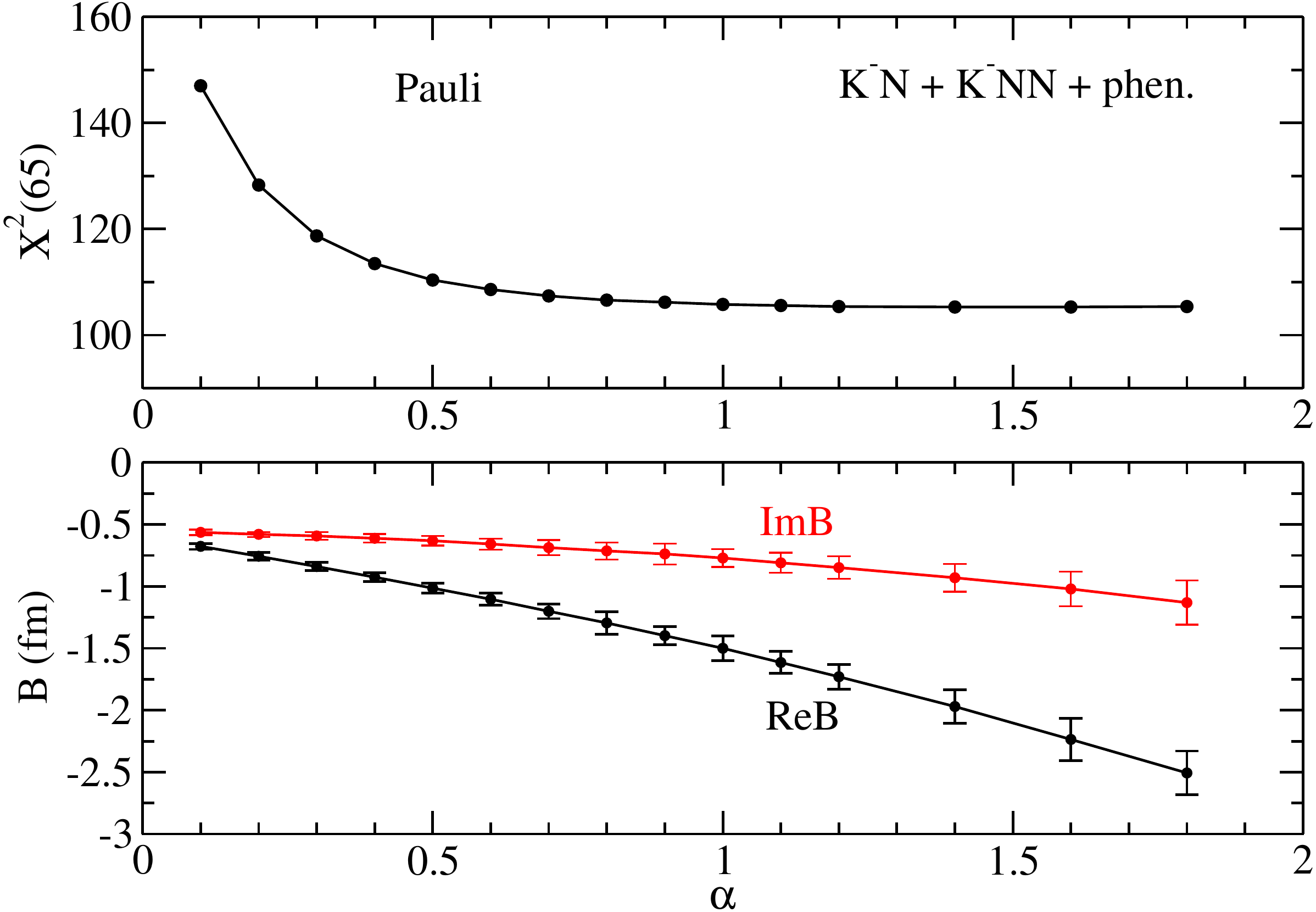}
\end{center}
\caption{Values of $\chi^2(65)$ (upper panel) and the complex amplitude $B$ (lower panel) as functions of parameter $\alpha$ obtained in the fit with $K^-N+K^-NN$ potentials based on the Pauli BCN amplitudes.}
\label{fig:fit_Pauli}
\end{figure}

Complete results of the fit, i.e. the values of $\chi^2(65)$, Re$B$, and Im$B$, scanned for values of $\alpha$ from 0 to 2 are presented in Figs.~\ref{fig:fit_WRW} and \ref{fig:fit_Pauli}. Fig. \ref{fig:fit_WRW} contains results obtained with $K^-$ potentials based on the WRW in-medium BCN amplitudes and Fig.~\ref{fig:fit_Pauli} those obtained with the Pauli BCN amplitudes. The Prague model, not presented here, yields similar results.
\begin{figure}[t]
\begin{center}
\includegraphics[width=0.9\textwidth]{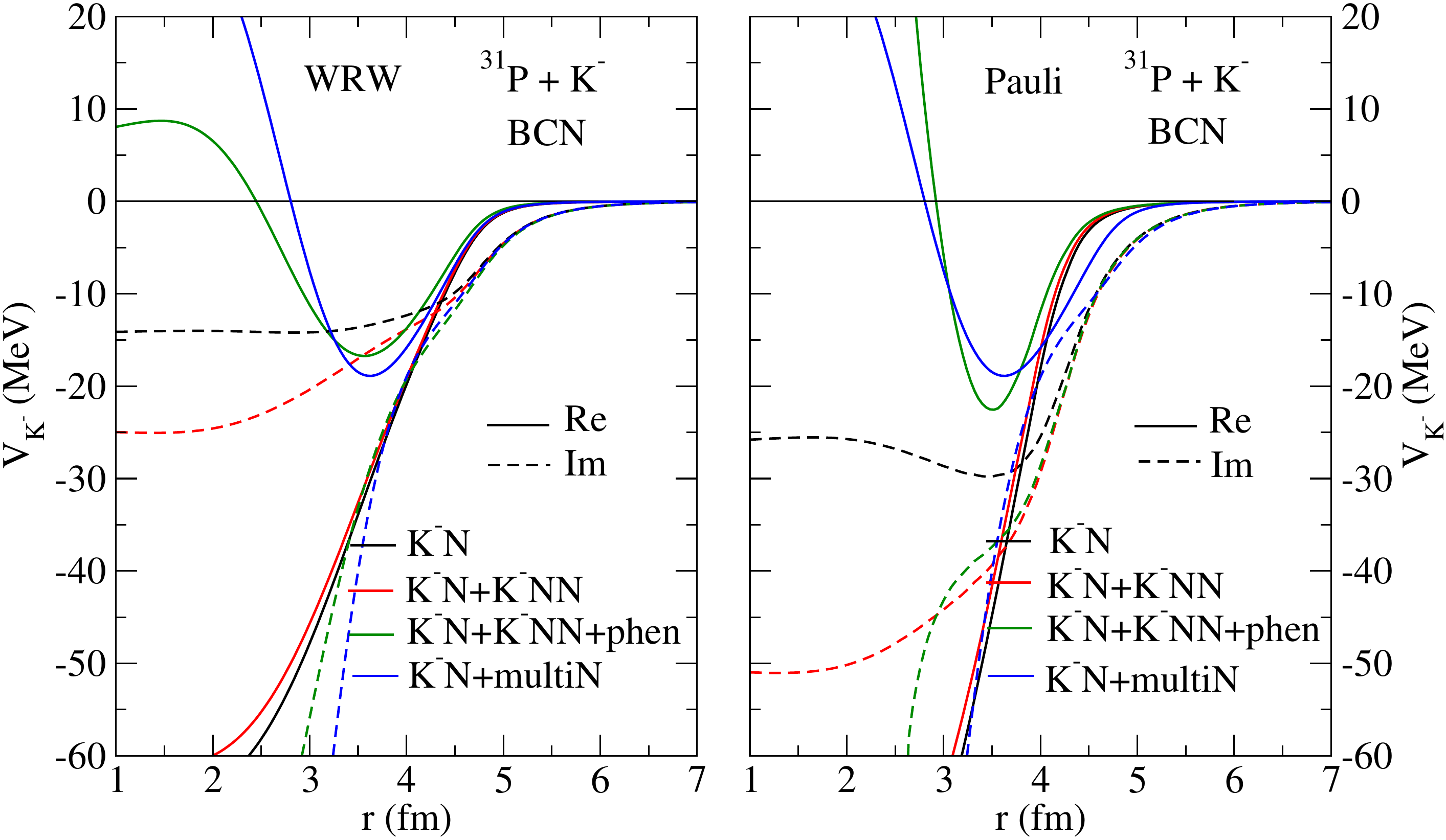}
\includegraphics[width=0.91\textwidth]{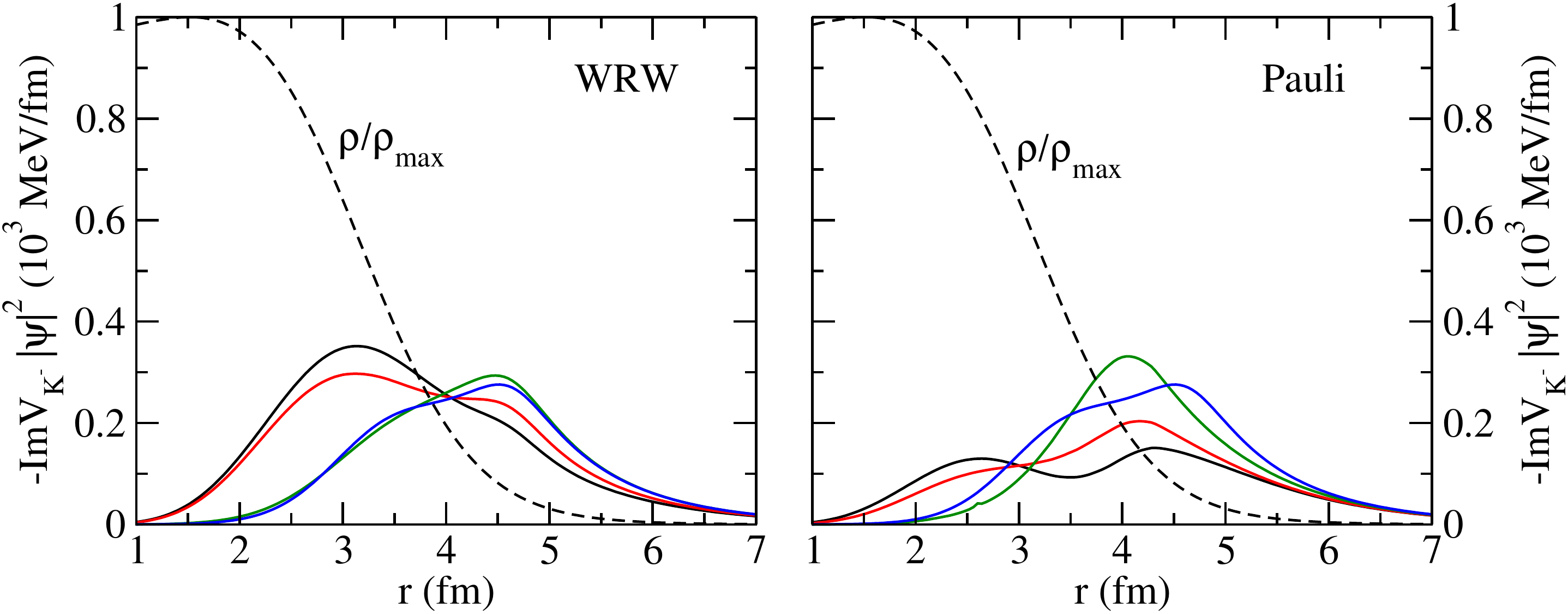}
\end{center}
\caption{Top panel presents total $K^-N$, $K^-N+K^-NN$, and $K^-N+K^-NN+$phen. potentials in $^{31}$P$+K^-$ calculated with the WRW (left panel) and Pauli (right panel) BCN amplitudes. The $K^-N$ + phenomenological multi-nucleon optical potential (denoted by '$K^-N+{\rm multi}N$') is also shown for comparison. Lower panel shows overlap of the $K^-$ wave function squared, $|\psi|^2$,  with Im$V_{K^-}$ for all considered potentials.}
\label{fig:31P+K}
\end{figure}

Top panel of Fig. \ref{fig:31P+K} demonstrates the depth of $K^-$ potentials in the $^{31}$P$+K^-$ atom calculated with the WRW (left panel) and Pauli (right panel) BCN amplitudes. In the lower panel, the overlaps of $K^-$ imaginary potential with $|\psi|^2$, where $\psi$ is the kaon radial wave function, are plotted for all considered potentials 
(note that the width of an atomic level is proportional to the overlap integral of Im$V_{K^-}$ and  $|\psi|^2$). 
The relative density $\rho/\rho_{max}$ is shown for illustration.  The pure $K^-$ single-nucleon potential (black lines) based on the WRW amplitudes differs considerably from the potential based on the Pauli amplitudes. Adding the microscopic $K^-NN$ potential (red lines) increases the depth of the imaginary potential in both cases and slightly decreases the depth of the real $K^-$ potential. The best fit $K^-N$+ phenomenological multi-nucleon potential ('$K^-N+{\rm multi}N$', blue lines) is presented for comparison. In the relevant density region, the $K^-N+K^-NN$ potential based on the Pauli amplitudes has considerably deeper imaginary part than the $K^-N$+ multi-N potential. This is compensated by  the negative value of Im$B$ for the additional phenomenological potential 
in the $K^-N+K^-NN +$phen. case (green line). The WRW modified amplitudes generate less absorptive $K^-N+K^-NN$ potential, leaving space for additional $3N (4N, ...)$ absorption. The overall $K^-N+K^-NN+$phen. potential evaluated using the WRW amplitudes is then very close to the $K^-N+$phen. multi-N potential in the relevant region. It is worth mentioning that the right balance between the real and imaginary parts of the $K^-$ potential is crucial. The $K^-N+$multi-N as well as $K^-N+K^-NN+$phen. potentials are less attractive, even repulsive, and very absorptive in the central region of the nucleus whereas the $K^-N$ and $K^-N+K^-NN$ potentials are strongly attractive and less absorptive inside the nucleus. As a consequence,  the overlap of Im$V_{K^-N}$ and Im$V_{K^-N+K^-NN}$ with the kaon wave function spans deeper inside the nucleus, towards higher densities, than in the $K^-N+K^-NN+$phen. and $K^-N+$multi-N cases.
\begin{table}[h!]
\caption{Values of $\chi^2$ for shifts $\Delta(\epsilon)$, widths $\Gamma$ for lower states and width $\Gamma^*$ for upper states in selected $K^-$ atoms, calculated with the $K^-N$, $K^-N+K^-NN$ and $K^-N+$phen. multi-N potentials based on the BCN Pauli and WRW modified amplitudes. Experimental values of $\Delta(\epsilon)$ and $\Gamma$ (in keV), and $\Gamma^*$ (in eV) including errors are shown in the last column for completeness.}
\vspace{10pt}
\label{Tab:chi2}
{\normalsize
 \begin{tabular}{c|c|c|c|c|c|c||c}
\multicolumn{2}{c|}{BCN} & \multicolumn{2}{|c|}{WRW} & \multicolumn{2}{|c|}{Pauli} & phen. & Exp.~\cite{fgbNPA94} \\ \hline
\multicolumn{2}{l|}{ ~~$\bm{\chi^2}$} & $K^-N$ & $+K^-NN$ & $K^-N$ & $+K^-NN$ & $K^-N$ + phen. multiN & \\ \hline \hline
& $\Delta(\epsilon)$ & 101.52  & 34.35 & 25.13 & 11.48 
& 1.76  & -0.59 (0.08)\\
$^{12}$C & $\Gamma$ & 44.80  & 27.45 & 17.00 & 9.44 
& 0.70  & ~1.73 (0.15)\\
& ~$\Gamma^*$ &1.71  & 1.47 & 0.15 & 0.67 
& 2.74 & ~0.99 (0.20)\\
 \hline
& $\Delta(\epsilon)$ & 41.04 & 15.13& 10.46 & 6.35 
& 0.03 & -0.33 (0.08)\\
$^{31}$P& $\Gamma$ & 13.72 & 10.34 & 11.43 & 6.42 
& 0.24 & ~1.44 (0.12)\\
& ~$\Gamma^*$ & 5.17 & 4.70 & 5.98 & 1.87 
& 0.30 & ~1.89 (0.30)\\
 \hline
 & {$\Delta(\epsilon)$} & {475.71} & {209.40} & {90.77} & {80.82} 
& {1.24}& {-0.494 (0.038)}\\
{$^{32}$S}& {$\Gamma$} & {0.76} & {2.83} &{67.35} & {43.29} 
& {9.24} & {~2.19 (0.10)}\\
& {~$\Gamma^*$} & {13.32} & {10.85} & {9.45} & {2.78} 
& {0.47} & {~3.03 (0.44)}\\
 \hline
  & $\Delta(\epsilon)$ & 38.27 & 17.69 & 4.23 & 4.62 
& 2.10 & -0.99 (0.17)\\
$^{35}$Cl& $\Gamma$ &5.94  & 2.56 & 10.94 & 5.39 
& 0.00 & ~2.91 (0.24)\\
& ~$\Gamma^*$ &7.92  &4.53  & 2.27 & 0.74 
& 0.15 & ~5.8 (1.70)\\
 \hline
   & $\Delta(\epsilon)$ & 33.50 & 8.93 & 1.54 & 2.71 
& 3.19 & -0.370 (0.047)\\
$^{63}$Cu& $\Gamma$ & 0.31  &0.02 & 4.90 & 3.57 
& 2.25 & ~1.37 (0.17)\\
& ~$\Gamma^*$ & 0.98 & 0.13 & 0.24 & 0.73 
& 1.52 & ~5.2 (1.1)\\
 \hline
    & $\Delta(\epsilon)$ & 9.00 &8.81  & 6.57 & 8.50 
& 2.15 & -0.41 (0.18)\\
$^{118}$Sn& $\Gamma$ &0.42  & 0.03 & 0.35 & 0.71
& 0.29 & ~3.18 (0.64)\\
& ~$\Gamma^*$ & 24.53 & 15.08 & 5.04 & 4.80 
& 4.09  & ~15.1 (4.4)\\
 \hline
     & $\Delta(\epsilon)$ & 7.52 & 3.67 & 3.24 & 4.84 
& 0.34 & -0.02 (0.012)\\
$^{208}$Pb& $\Gamma$ & 0.12 & 0.10 & 0.31 & 0.38
& 0.39 & ~0.37 (0.15)\\
& ~$\Gamma^*$ & 0.06 & 0.18 & 0.35 & 0.41 
& 0.52  & ~4.1 (2)\\
 \hline
 \hline
$\bm{\chi^2}$& \textbf{total} & \bf{820.37} & \bf{378.24} & \bf{277.69} & \bf{200.54}
& \bf{33.71} & \\
& \textbf{$^{32}$S out} & \bf{330.58} & \bf{155.16} & \bf{110.13} & \bf{73.65}
& \bf{22.76}  & \\
\end{tabular}}
\end{table} 

In Table \ref{Tab:chi2}, we present values of $\chi^2$ for shifts and widths for lower state and width for upper state in selected $K^-$ atoms (21 data points), calculated using the $K^-N$, $K^-N+K^-NN$ and $K^-N+$phen. multi-N potentials based on the Pauli and WRW modified BCN amplitudes. 
When the microscopic $K^-NN$ absorption is taken into account the description of the atomic data improves considerably -- $\chi^2 (21)$ decreases from $\approx 800$ to $\approx 400$ for the WRW amplitudes and from $\approx 300$ to $\approx 200$ for the Pauli amplitudes. 
Note that the most pronounced contribution to the total  $\chi^2 (21)$ comes from $^{32}$S due to a very small experimental error in energy shift. This may imply that the microscopic potentials are not able to describe the energy shift reasonably well. However, the experimental data on $\Delta(\epsilon)$ and $\Gamma$ (in keV), and $\Gamma^*$ (in eV) presented with corresponding errors in the last column of Table \ref{Tab:chi2} for completeness, were compiled by Batty (private communication) and ~\cite{fgbNPA94} as weighted averages. In the case of sulfur, the data come from three different experiments (see Ref.~\cite{fgbNPA94}) and the values of energy shifts and widths span over a quite large range. For instance, if only the data from Ref.~\cite{backenstossPLB72} are considered, the total value of $\chi^2$ for sulfur drops down significantly, e.g., from $\chi^2 = 127$ to $\chi^2 = 42$ for $K^-N+K^-NN$ Pauli potentials! Moreover, when the $^{32}$S data are excluded from the fit, the total $\chi^2$ drops by more than a half (see the last two rows in Table \ref{Tab:chi2}). This indicates that a new measurement of kaonic sulfur is needed. 
It would be desirable to remeasure kaonic sulfur within current experiments. 

Next, we calculated branching ratios for $K^-N$ and $K^-NN$ absorption channels in $^{12}$C$+K^-$ using the WRW and Pauli BCN and P amplitudes. The branching ratios were evaluated as fractions of partial width in the respective channel (see Table~\ref{tab:channels}) over the total width
\begin{equation}
\text{BR} = \frac{\Gamma_{\rm channel}}{\Gamma_{\rm total}} = \frac{\int \text{Im}V_{\rm channel}(r) |\psi(r)|^2 dr }{\int \text{Im}V_{K^-}(r)|\psi(r)|^2 dr }~,    
\end{equation}
where $\psi(r)$ is the $K^-$ radial wave function.
The branching ratios for the lower (l=1) and upper (l=2) states are presented in Table~\ref{tab:ratios1} and Table~\ref{tab:ratios2}, respectively. 
\begin{table}[b!]
\caption{Primary-interaction branching ratios (in $\%$) for mesonic ($K^-N\rightarrow Y \pi$, $Y=\Lambda, \Sigma$) and non-mesonic ($K^-NN\rightarrow Y N$) absorption  of $K^-$ from the l=1 state in $^{12}$C$+K^-$, calculated with $K^-N+K^-NN$ potentials based on the WRW and Pauli BCN and P amplitudes. The experimental data for primary-interaction branching ratios are shown for comparison.} 
\label{tab:ratios1}
{\normalsize
 \begin{tabular}{l|c|c||c|c||c}
 $^{12}$C + $K^-$ (l=1)  & \multicolumn{2}{c||}{BCN}  & \multicolumn{2}{c||}{P} & Exp.~\cite{bubble3}  \\ \hline
mesonic ratio  & ~WRW~ & ~Pauli~  & ~WRW~ & ~Pauli~ & $^{12}$C \\ \hline \hline
$\Sigma^+ \pi^- $ & 25.3 & 21.7  & 24.5 &20.0 & 29.4 $\pm$ 1.0\\
$\Sigma^- \pi^0$ & 7.7 & 6.7 & 7.3 &5.3 & 2.6 $\pm$ 0.6\\
$\Sigma^- \pi^+$ & 7.6 & 12.9 &7.7 &14.7 & 13.1 $\pm$ 0.4\\
$\Sigma^0 \pi^-$ &7.8 &6.8 &7.4 & 5.3& 2.6 $\pm$ 0.6\\
$\Sigma^0 \pi^0 $ & 12.7 & 14.2 &12.5 &15.0 & 20.0 $\pm$ 0.7\\
$\Lambda \pi^0$  & 6.0 & 5.0 &5.2 &3.8 & 3.4 $\pm$ 0.2\\ 
$\Lambda \pi^-$ & 11.8 & 10.1 &10.3 &7.4 & 6.8 $\pm$ 0.3\\ \hline
total 1N ratio& 79.0& 77.4 &75.6 &74.6 & 77.9 $\pm$ 1.6\\ \hline 
$R_{\pm}=\frac{(\Sigma^+ \pi^-)}{(\Sigma^- \pi^+)}$ &3.3 & 1.7& 3.2 & 1.4 & 2.24 $\pm$ 0.12 \\ 
$R_{pn}=\frac{(\Sigma^+ \pi^-)+(\Sigma^- \pi^+)}{(\Sigma^- \pi^0)}$ &4.3 &5.2 &4.4 &6.6  & 16.3 $\pm$ 4.0  \\ \hline \hline
non-mesonic ratio & ~WRW~ & ~Pauli~ &  ~WRW~ & ~Pauli~ & 76\% CF$_3$Br + 24\% C$_3$H$_8$~\cite{bubble1}  \\ \hline 
$\Lambda p + \Lambda n +\Sigma^0 p + \Sigma^0 n$ & 11.4 & 11.8  & 12.8& 13.3&  14.1 $\pm$ 2.5$~^{\rm{a}}$  \\
$\Sigma^- p + \Sigma^- n$  & 3.8 & 4.4  &5.0 & 5.2& 7.3 $\pm$ 1.3$~^{\rm{a}}$ \\
$\Sigma^+ n $  & 5.7 & 6.4 &6.6 & 6.9& 4.3 $\pm$ 1.2$~^{\rm{a}}$  \\
$\Sigma^0 p + \Sigma^0 n$ & 4.6 &  5.2  & 5.6 & 5.8&  -  \\ \hline 
\multirow{2}*{total 2N ratio}  & \multirow{2}*{21.0} & \multirow{2}*{22.6}  & \multirow{2}*{24.4} & \multirow{2}*{25.4} & 25.7 $\pm$ 3.1~\cite{bubble1} \footnote{multinucleon capture rate} \\
&&&&& $16\pm 3(\text{stat.})^{+4}_{-5}(\text{syst.})$~\cite{amadeus19}
\end{tabular}}
 \end{table}
\begin{table}[t!]
\caption{Primary-interaction branching ratios (in $\%$) for mesonic ($K^-N\rightarrow Y \pi$, $Y=\Lambda, \Sigma$) and non-mesonic ($K^-NN\rightarrow Y N$) absorption of $K^-$ from the l=2 state in $^{12}$C$+K^-$, calculated with $K^-N+K^-NN$ potentials based on the WRW and Pauli blocked BCN and P amplitudes. The experimental data for primary- interaction branching ratios are shown for comparison.}
\label{tab:ratios2}
{\normalsize
 \begin{tabular}{l|c|c||c|c||c}
$^{12}$C + $K^-$ (l=2) & \multicolumn{2}{c||}{BCN} & \multicolumn{2}{c||}{P} & Exp.~\cite{bubble3}  \\ \hline
mesonic ratio  & ~WRW~ & ~Pauli~  & ~WRW~ & ~Pauli~ & $^{12}$C \\ \hline \hline
$\Sigma^+ \pi^- $ & 26.9 & 22.4 &28.1 &22.1 & 29.4 $\pm$ 1.0\\
$\Sigma^- \pi^0 $ & 8.3 & 7.7 &7.2 &5.9 & 2.6 $\pm$ 0.6\\
$\Sigma^- \pi^+ $ & 15.5 & 17.5 & 17.1&17.6 & 13.1 $\pm$ 0.4\\
$\Sigma^0 \pi^-$ &8.4 &7.9 &7.3 &5.9 & 2.6 $\pm$ 0.6\\
$\Sigma^0 \pi^0 $ & 17.2 & 16.4 &19.3 &17.3 & 20.0 $\pm$ 0.7\\
$\Lambda \pi^0 $  & 5.2 & 5.0 &4.2 &3.7 & 3.4 $\pm$ 0.2\\ 
$\Lambda \pi^-$ & 10.4 & 9.9 &8.3 &7.2  & 6.8 $\pm$ 0.3\\ \hline
total 1N ratio& 91.9& 87.0 &90.7 &82.0 & 77.9 $\pm$ 1.6\\ \hline 
$R_{\pm}=\frac{(\Sigma^+ \pi^-)}{(\Sigma^- \pi^+)}$ &1.7 & 1.3&1.6 &1.3 & 2.24 $\pm$ 0.12 \\ 
$R_{pn}=\frac{(\Sigma^+ \pi^-)+(\Sigma^- \pi^+)}{(\Sigma^- \pi^0)}$ &5.1 &5.2 &6.3 &6.7 & 16.3 $\pm$ 4.0  \\ \hline \hline
non-mesonic ratio & ~WRW~ & ~Pauli~ & ~WRW~ & ~Pauli~ & 76\% CF$_3$Br + 24\% C$_3$H$_8$~\cite{bubble1}  \\ \hline 
$\Lambda p + \Lambda n +\Sigma^0 p + \Sigma^0 n$ & 4.2 & 6.7  &4.6 &9.0 &  14.1 $\pm$ 2.5$~^{\rm{a}}$  \\
$\Sigma^- p + \Sigma^- n$  & 1.7 & 3.1  &2.1 &4.2 &7.3 $\pm$ 1.3$~^{\rm{a}}$ \\
$\Sigma^+ n $  & 2.2 & 3.5 &2.6 &4.8 & 4.3 $\pm$ 1.2$~^{\rm{a}}$  \\
$\Sigma^0 p + \Sigma^0 n$ & 1.9 &  3.1  &2.2 & 4.2&  -  \\ \hline 
\multirow{2}*{total 2N ratio}  & \multirow{2}*{8.1} & \multirow{2}*{13.0} & \multirow{2}*{9.3} & \multirow{2}*{18.0} & 25.7 $\pm$ 3.1~\cite{bubble1} \footnote{multinucleon capture rate} \\
&&&&& $16\pm 3(\text{stat.})^{+4}_{-5}(\text{syst.})$~\cite{amadeus19}
\end{tabular}}
\end{table} 
Theoretical values are compared with experimental data on primary-interaction branching ratios, i.e., corrected for secondary interactions of particles created in the absorbing nucleus, measured in old bubble chamber experiments \cite{bubble3, bubble1}. 
Note that the branching ratios are evaluated only for $K^-$ single-nucleon and $K^-$ two-nucleon absorptive potentials and the effect of $3N(4N)$ absorption is not taken into account. Both BCN and P models yield the values of branching ratios in reasonable agreement with experimental data, except the  branching ratios for $\Sigma^-\pi^0$ and $\Sigma^0\pi^-$ production, which differ from the experimental values for all considered amplitudes.
The WRW and Pauli in-medium amplitudes in both interaction models yield comparable branching ratios for the total 1N and 2N absorption in the l=1 case (see Table~\ref{tab:ratios1}). In the l=2 case (see Table~\ref{tab:ratios2}), there are evident differences between the total branching ratios calculated using various  in-medium amplitudes under consideration.

Valuable information about absorption of an antikaon in the nuclear medium has been provided recently by the AMADEUS Collaboration which measured branching fractions for the $K^-$ two-nucleon absorption in reactions of low-energy $K^-$ with a carbon target \cite{amadeus16, amadeus19}. The total 2N absorption ratios calculated within the BCN model 
(presented in Tables~\ref{tab:ratios1} and \ref{tab:ratios2}) could be considered consistent with the value measured by the AMADEUS Collaboration, BR($K^-$2N$\rightarrow$YN) = $(16\pm 3(\text{stat.})^{+4}_{-5}(\text{syst.}))\% $ \cite{amadeus19}. The P model yields branching ratios for the $K^-$ two-nucleon absorption from the l=1 state slightly above the error bars.

\begin{table}[b]
\caption{Branching ratio (in \%) for $\Lambda N$ and $\Sigma^0 N$ production in $K^-NN$ absorption at rest ($p_{K^-}=0$~MeV/c) in the $^{12}$C$+ K^-$ atom for the lower (l=1) and upper (l=2) state, calculated with $K^-N + K^-NN$ potentials based on the WRW modified and Pauli amplitudes derived from the BCN and P models. Theoretical values are compared with the AMADEUS data. }
\label{tab:ratios_LN_S0N}
\begin{tabular}{l|c|c|c|c||c|c|c|c|c||c}
\textbf{BCN}~&\multicolumn{2}{c|}{WRW} & \multicolumn{2}{c||}{Pauli} & ~\textbf{P}~ &\multicolumn{2}{c|}{WRW}& \multicolumn{2}{c||}{Pauli} & \\ \hline
BR~~  & ~l=1~ & ~l=2~ & ~l=1~ & ~l=2~ & &~l=1~ & ~l=2~ & ~l=1~ & ~l=2~ &  Exp.~\cite{amadeus19} \\ \hline \hline
$\Lambda N$ & 5.45 & 2.32 & 4.23 & 3.02 & &5.27 & 2.17&4.30 &3.19 & 6.45 $\pm~ 1.41$(stat.)$^{+0.5}_{-0.6}$(syst.) \\ \hline
$\Sigma^0N$ &  4.44 & 2.09 & 3.99 &2.93 & &5.11 &2.37 &4.53 &3.47 & 7.55 $\pm~ 2.2$(stat.)$^{+4.2}_{-5.4}$(syst.)
\end{tabular}
\end{table}

The AMADEUS Collaboration determined among others branching ratios for quasi-free (QF) production of $\Lambda p$ and $\Sigma ^0 p$ pairs from the $K^-$ two-nucleon absorption in $^{12}$C, without final state interaction (FSI) with a residual nucleus, and  branching ratios for processes where the primary created $\Lambda(\Sigma^0)$'s  undergo elastic FSI (see Table 1 in Ref.~\cite{amadeus19}). The FSI ratios include also channels $K^-pn \rightarrow \Lambda n$ and $K^-pn \rightarrow \Sigma^0 n$. It is to be noted that the AMADEUS data do not include QF production of $\Lambda n$ and $\Sigma^0 n$ pairs.  In our present calculations of kaonic atoms, we evaluated branching ratios for the total $\Lambda N$ and $\Sigma^0 N$ production. The branching ratios calculated within our microscopic model  are compared with experimental data on the QF+FSI branching ratios in Table~\ref{tab:ratios_LN_S0N}. The theoretical values of branching ratios for both $\Lambda N$ and $\Sigma^0 N$ production in the $K^- NN$ absorption from the l=1 state in the $^{12}{\rm C} + K$ atom are in agreement with the experimental data for both the BCN and P model. 
As for the l=2 state, the branching ratios for $\Lambda N$ and $\Sigma^0 N$ production
are significantly lower than the ratios for $K^- NN$ absorption from the l=1 state. 
Nonetheless, the calculated $\Sigma^0 N$ branching ratios could be still considered consistent with experiment due to rather large experimental errors. 
Following the finding that 75\% of $K^-$ absorption in $^{12}$C takes place from the upper $l=2$ state~\cite{fgNPA17}, we evaluated weighted average of the l=1 (25 \%) and l=2 (75\%) values of the $\Lambda N$ branching ratios within the BCN model for completeness. The resulting values BR($\Lambda N)_{\rm WRW}$~=~3.1~\% and BR($\Lambda N)_{\rm Pauli}$~= 3.3~\% are well below the experimental value and again out of the experimental error. 

Finally, the AMADEUS Collaboration reported the ratio of branching ratios~\cite{amadeus19}
\begin{equation}
    R=\frac{{\rm BR}(K^-pp\rightarrow \Lambda p)}{{\rm BR}(K^-pp \rightarrow \Sigma^0 p)}=0.7 \pm~ 0.2(\text{stat.})^{+0.2}_{-0.3}(\text{syst.})~.
\end{equation}
We calculated this ratio in $^{12}$C$+ K^-$ atom, using the $K^-N+K^-NN$ microscopic potentials based on the BCN and P amplitudes. The value of $R$ is about 1 for both the WRW and Pauli in-medium amplitudes and both interaction models, which is within error bars of the experimental value, as demonstrated in Table~\ref{tab:ratio_R}. 
 
\begin{table}[h!]
\caption{Ratio $R$ in the $^{12}$C$+ K^-$ atom for lower (l=1) and upper (l=2) state and $K^-NN$ absorption at rest ($p_{K^-}=0$~MeV/c), calculated with $K^-N + K^-NN$ potentials based on the WRW modified and Pauli amplitudes from the BCN and P models. Theoretical values are compared with the AMADEUS data. }
\label{tab:ratio_R}
\begin{tabular}{l|c|c|c|c|c}
&\multicolumn{2}{c|}{WRW} & \multicolumn{2}{c|}{Pauli} & \\ \hline
$R$~ & ~l=1~ & ~l=2~ & ~l=1~ & ~l=2~ & Exp.~\cite{amadeus19} \\ \hline 
BCN & 1.1 & 1.0 & 1.1 & 1.0 & \multirow{2}{*}{0.7 $\pm~ 0.2$(stat.)$^{+0.2}_{-0.3}$(syst.)} \\
P &  1.1 & 1.0 & 1.0 & 1.0 & 
\end{tabular}
\end{table}

\section{Conclusions}
\label{conclusions}
In this work, we calculated strong-interaction energy shifts and widths in  various kaonic atoms, from lithium up to uranium. For the first time, the calculations were performed using microscopic $K^-N+K^-NN$ potentials based on scattering  amplitudes derived from two chiral coupled-channels meson-baryon interaction models -- the Barcelona and Prague model. 

The $K^-NN$ potentials were constructed within our recently formulated microscopic $K^-NN$ absorption model \cite{hrPRC20}. We took into account medium modifications of the free-space amplitudes due to the Pauli principle. They were incorporated by two different methods -- i) Pauli blocking included directly in the chiral amplitudes and ii) WRW procedure based on multiple scattering approach. 

The $K^-N+K^-NN$ potentials, based on in-medium chiral amplitudes, calculated for 23 nuclear species were confronted with kaonic atom data. The value of $\chi^2(65)$ significantly improves when the $K^-$ two-nucleon potentials are included for both the Pauli and WRW in-medium amplitudes. It drops approximately to one half of the value corresponding to just $K^-N$ chiral potential, nonetheless, it remains still sizable. Next, we added a phenomenological term to the microscopic $K^-N+K^-NN$ potentials in order to incorporate and quantify missing $K^{-}-3N(4N)$ processes. After adding this term the description of the data further improved and the resulting $\chi^2(65) \approx 100$ was comparable with the best fit $K^-N$ + phenomenological multi-nucleon potential. However, the Pauli in-medium amplitudes yielded negative imaginary amplitude $B$ for the additional phenomenological term. This implies that the $K^-N+K^-NN$ potentials based on the Pauli amplitudes are too absorptive and there is no space for additional absorption from $3N(4N)$ processes, which is in contrast to experimental measurements of the AMADEUS Collaboration~\cite{amadeus19}. 

Unanticipated values of the fitted parameters of the additional phenomenological term indicate certain deficiencies in microscopic potentials in the region where $K^-$ absorption takes place, i.e., in their density dependence. Our microscopic model could be further improved by introducing hadron self-energies as another component of in-medium modifications of $K^-N$ scattering amplitudes in the construction of $K^-N$ and $K^-NN$ potentials.    
This extension of our microscopic model is currently under investigation and will be published elsewhere. 

On the other hand, the analyses of kaonic atom data including microscopic as well as phenomenological $K^-$ optical potentials hint at certain  inconsistencies in the available experimental data which were compiled from different experiments, performed in the 1970s. This indicates that it would be opportune to repeat some of these measurements. 

Finally, we calculated branching ratios for all $K^-N$ and $K^-NN$ absorption channels in the $^{12}$C$+K^-$ atom, using microscopic $K^-N+K^-NN$ potentials based on the in-medium BCN and P amplitudes. Our results are in reasonable agreement with old bubble chamber data on primary-interaction branching ratios. The total ratio for the $K^-$ two-nucleon absorption was found to be in accordance with the latest measurement by the AMADEUS Collaboration, BR($K^-$2N$\rightarrow$YN) = $(16\pm 3(\text{stat.})^{+4}_{-5}(\text{syst.}))\% $. Moreover, the newly measured ratio $R=({\rm BR}(K^-pp\rightarrow \Lambda p))/{\rm BR}(K^-pp \rightarrow \Sigma^0 p))$ was found to be consistent with the one calculated in this work. The AMADEUS Collaboration reported also branching ratios for the $\Lambda N$ and $\Sigma^0N$ production in $K^-$ two-nucleon absorption. Here, only the results for the $K^-$ absorption from the lower l=1 state are in agreement with the data, while the branching ratios for $K^-$ absorption from the upper l=2 state are too low. This is in contradiction with the empirical fact that $75\%$ of $K^-$ absorption takes place from the upper level in $^{12}$C.

To summarize, although some positive results have been presented in this work,  confrontation of our microscopic model with all available kaonic atom data revealed need for a further theoretical as well as experimental study in order to get a consistent description of $K^-$ absorption in the nuclear medium.

\section*{Acknowledgements}
We thank \`{A}ngels Ramos for providing us with the BCN model amplitudes and for careful reading of the manuscript.
J.O. and J.M. acknowledge support from the Czech Science Foundation GACR Grant No. 19-19640S. 
The present work is part of a project funded by the European Union's
            Horizon 2020 research \& innovation programme, Grant Agreement No. 824093

\end{document}